\documentclass[12pt]{article}
\begin{document}
\begin{center}{\Large \bf
Empirical determinations of Feynman integrals\\using integer relation algorithms}\\[4pt]
{\large Kevin Acres\footnote{Cryptography Research Centre, 
Technology Innovation Institute, Abu Dhabi, UAE,\\{\tt kevin@tii.ae}}
and David Broadhurst\footnote{School of Physical Sciences, Open University, 
Milton Keynes MK7 6AA, UK,\\{\tt david.broadhurst@open.ac.uk}}, 10 March 2021}\end{center}
{\bf Abstract}: Integer relation algorithms can convert numerical results
for Feynman integrals to exact evaluations, when one has reason to suspect
the existence of reductions to linear combinations of a basis,
with rational or algebraic coefficients.
Once a tentative reduction is obtained, confidence
in its validity is greatly increased by computing more decimal digits of the
terms and verifying the stability of the result. Here we give examples
of how the PSLQ and LLL algorithms have yielded
remarkable reductions of Feynman integrals to multiple polylogarithms
and to the periods and quasi-periods of modular forms. Moreover,
these algorithms have revealed quadratic relations between Feynman
integrals. A recent application concerning black holes involves quadratic
relations between combinations of Feynman integrals with algebraic coefficients.

\section{Introduction}

The mathematical problem at hand is easy to state: given numerical approximations to $n>2$ real numbers, $x_k$, is
there at least one {\em probable} relation $\sum_{k=1}^{n}z_k x_k =0$
with integer coefficients $z_k$, at least two of which are non-zero? If so, produce such a relation.

By way of example, in 1985 Broadhurst studied periods coming from 6-loop counterterms~\cite{5lp}
in $\phi^4$  theory and found, with good confidence, the relations
\begin{equation}
P_{6,1}=168\zeta_9,\quad P_{6,2}=\frac{1063}{9}\zeta_9+8\zeta_3^3,\quad 16P_{6,3}+P_{6,4}=1440\zeta_5\zeta_3
\label{P6}\end{equation}
between the periods $P_{6,k}$, as labeled in the later census by Schnetz~\cite{census}, and
Riemann zeta values $\zeta_s=\sum_{n>0}1/n^s$. There was a strong intuition that $P_{6,3}$ and $P_{6,4}$
might involve the multiple zeta value (MZV)
\begin{equation}
\zeta_{5,3}=\sum_{m>n>0}\frac{1}{m^5n^3}=0.03770767298484754401130478\ldots
\label{Z53}\end{equation}
along with $\zeta_8$ and the product $\zeta_5\zeta_3$. Yet such relations were not discovered, at the
level of accuracy then attainable.

A decade later, Broadhurst and Kreimer~\cite{bk} solved this problem, by improving accuracy
for the periods and using the PSLQ algorithm developed by Ferguson and Bailey~\cite{FB}, which 
identified
\begin{equation}
P_{6,3}=\mbox{$\frac{36}{5}$}\left(12\zeta_{5,3}-29\zeta_8\right)+252\zeta_5\zeta_3.\label{P63}
\end{equation}
Moreover, they found $\zeta_{3,5,3}$, with weight 11 and depth 3,  in some 7-loop periods.

Much experimenting with PSLQ led to the Broadhurst-Kreimer conjecture~\cite{bkc} that the number 
$N(w,d)$ of {\em primitive} MZVs of weight $w$ and depth $d$ is generated by
\begin{equation}\prod_{w>2}\prod_{d>0}(1-x^w y^d)^{N(w,d)}
=1-\frac{x^3y}{1-x^2} +
\frac{x^{12}y^2(1-y^2)}{(1-x^4)(1-x^6)}
\label{BK}\end{equation}
with a final term inferred by relating MZVs to {\em alternating} sums.

\section{PSLQ and LLL}

PSLQ came from work by Helaman Ferguson and Rodney Forcade~\cite{FF} in 1977,
implemented in {\em multiple-precision} {\sc fortran} by David Bailey in 1992,
improved and parallelized by Bailey and Broadhurst~\cite{BB} in 1999,
and used extensively in the study of Feynman integrals since then.

The algorithm proceeds as follows. First we {\em initialize}:
 \begin{enumerate}
\item For $j := 1$ to $n$: for $i := 1$ to $n$: if $i = j$ then set
$A_{i j} := 1$ and $B_{i j} := 1$ else set $A_{i j} := 0$ and $B_{i j}
:= 0$; endfor; endfor.
\item For $k := 1$ to $n$: set $s_k := {\bf sqrt}\left(\sum_{j=k}^n x_j^2\right)$;
endfor.  Set $t = 1 / s_1$.\\  For $k := 1$ to $n$: set $y_k := t x_k; \;
s_k := t s_k$; endfor.
\item For $j := 1$ to $n - 1$: for $i := 1$ to $j - 1$:
set $H_{i j} := 0$; endfor;\\ set $H_{j j} := s_{j+1} / s_j$; for $i :=
j + 1$ to $n$: set $H_{i j} := - y_i y_j / (s_j s_{j+1})$; endfor;
endfor.
\item For $i := 2$ to $n$: for $j := i- 1$ to $1$ step
$-1$: set $t := {\bf round} (H_{i j} / H_{j j})$; $y_j := y_j + t y_i$;
for $k := 1$ to $j$: set $H_{i k} := H_{i k} - t H_{j k}$; endfor; \\ for
$k := 1$ to $n$: set $A_{i k} := A_{i k} - t A_{j k}$,  $B_{k j} :=
B_{k j} + t B_{k i}$; endfor; endfor; endfor.
\end{enumerate}
Thus the numerical data, in the vector $x$, is converted to a vector $y$, by taking
square roots of partial sums of squares. Then the matrices $H$, $A$ and $B$ are created,
with a crucial {\em rounding} in Step 4.  

Then we proceed by {\em iteration}:
\begin{enumerate}
\item Select $m$ such that $({\bf 4/3})^{i/2} |H_{i i}|$ is maximal when $i = m$.
{\bf Swap} the entries of $y$ indexed $m$ and $m + 1$, the
corresponding rows of $A$ and $H$, and the corresponding columns of $B$.
\item If $m \leq n - 2$ then set $t_0
:= {\bf sqrt}(H_{m m}^2 + H_{m,m+1}^2)$, $t_1 := H_{m m} / t_0$ and $t_2 :=
H_{m,m+1} / t_0$; for $i := m$ to $n$: set $t_3 := H_{i m}$, $t_4 :=
H_{i, m+1}$, $H_{i m} := t_1 t_3 + t_2 t_4$ and $H_{i, m+1} := - t_2
t_3 + t_1 t_4$; endfor; endif.
\item For $i := m + 1$ to $n$: for $j := \min(i - 1, m +
1)$ to $1$ step $-1$: set $t := {\bf round} (H_{i j} / H_{j j})$ and $y_j :=
y_j + t y_i$; for $k := 1$ to $j$: set $H_{i k} := H_{i k} - t H_{j
k}$; endfor; for $k := 1$ to $n$: set $A_{i k} := A_{i k} - t A_{j k}$
and $B_{k j} := B_{k j} + t B_{k i}$; endfor; endfor; endfor.
\item If the largest entry of $A$ exceeds the precision, then {\bf fail}, else
if a component of the $y$ vector is very small,
then output the {\bf relation} from the corresponding column of $B$, 
else go back to Step 1.
\end{enumerate}
The constant $4/3$ in the {\em swap} of Step 1 of the iteration ensures that the algorithm will
find a relation, provided that one exists and the data have been specified to sufficient accuracy.
In practice, reliable results may be obtained with a smaller constant.
 
For big problems, {\em parallelized} PSLQ~\cite{BB} has been vital, especially
for the magnetic moment of the electron~\cite{SL}. For smaller problems, there
is an alternative.

\subsection{LLL}

In 1982, Arjen Lenstra, Hendrik Lenstra and L\'{a}szl\'{o} Lov\'{a}sz gave 
the LLL algorithm~\cite{LLL} for lattice reduction to a basis with short and almost orthogonal components.
An extension of this underlies the {\tt lindep} procedure in Pari-GP~\cite{GP}, which we here apply
to the problem of determining the period $P_{6,3}$.
\begin{verbatim}
P63=107.71102484102;\\ only 14 digits specified
V=[P63,zetamult([5,3]),zeta(8),zeta(5)*zeta(3)];
for(d=11,16,U=lindep(V,d);U*=sign(U[1]);print([d,U~]));
[11, [4, -827, 173, -460]]
[12, [4, -827, 173, -460]]
[13, [4, -827, 173, -460]]
[14, [5, -432, 1044, -1260]]
[15, [5, -432, 1044, -1260]]
[16, [196, 1652, -9045, -9701]]
\end{verbatim}

In this case, using 14 good digits of $P_{6,3}$, we happen to obtain the correct result~(\ref{P63}).
In practice, one would need several more digits, for better confidence. In what follows,
all claimed relations have been checked using at least 100 more digits than
were needed for the discoveries, making the probability of a mistake less than $1/10^{100}$.
 
\subsection{Improvement and parallelization of PSLQ}

{\bf Multi-level} improvement: perform most operations at 64-bit precision, some at intermediate
precision (Bailey and Broadhurst~\cite{BB} chose 125 digits) and only the bare minimum 
of the most delicate operations
at  full precision (more than 10000 digits, for some big problems).

{\bf Multi-pair} improvement: swap up to $0.4n$ disjoint pairs of the $n$ indices at each iteration.
In this case, it is not proven that the algorithm will succeed, but it has not yet been found to fail.

{\bf Parallelization:} distribute the disjoint-pair jobs; for each pair, distribute
the full-precision matrix multiplication in the outermost loop.

\subsection{Examples}

Bailey and Broadhurst~\cite{BB}, working at 10000 digits, found that the constant associated
with the fourth bifurcation of the logistic map is the root of a polynomial of degree 240.

They tested a conjecture on alternating sums of the form
\begin{equation}
\zeta\left(\begin{array}{cccc}
\sigma_1,&\sigma_2&\cdots&\sigma_d\\
s_1,&s_2&\cdots&s_d\end{array}\right)=
\sum_{k_1 > k_2 > \cdots > k_d > 0}
\frac{\sigma_1^{k_1}}{k_1^{s_1}}\,
\frac{\sigma_2^{k_2}}{k_2^{s_2}}\,\cdots\,
\frac{\sigma_d^{k_d}}{k_d^{s_d}}
\label{nest} \end{equation}
where $\sigma_j=\pm1$ are signs and $s_j>0$ are integers,
namely that at weight $w=\sum_j s_j$ every convergent alternating sum
is a rational linear combination of elements of a basis of size 
$F_{w+1}=F_w+F_{w-1}$, i.e.\ the Fibonacci number with index $w+1$.
At $w=11$, many integer relations of size
$F_{12}+1=145$ were found, at 5000-digit precision.

For weights $w\le20$, inverse binomial sums~\cite{binom} of the form
\begin{equation}S(w)=\sum_{n=1}^\infty\frac{1}{n^w{2n\choose n}}
\label{IB}\end{equation}
were reduced to multiple polylogarithms of the sixth root of unity~\cite{sixth}, 
with $S(20)$ given by 106 terms.

\subsection{Relations in the multiple zeta value data mine}

The Broadhurst-Kreimer conjecture~(\ref{BK}) came from the PSLQ discovery that
\begin{eqnarray}
2^5\cdot3^3\zeta_{4,4,2,2}&=&2^{14}\sum_{m>n>0}\frac{(-1)^{m+n}}{(m^3n)^3}
+2^5\cdot3^2\,\zeta_3^4
+2^6\cdot3^3\cdot5\cdot13\,\zeta_9\,\zeta_3\nonumber\\&&{}
+2^6\cdot3^3\cdot7\cdot13\,\zeta_7\,\zeta_5
+2^7\cdot3^5\,\zeta_7\,\zeta_3\,\zeta_2
+2^6\cdot3^5\,\zeta_5^2\,\zeta_2\nonumber\\&&{}
-2^6\cdot3^3\cdot5\cdot7\,\zeta_5\,\zeta_4\,\zeta_3
-2^8\cdot3^2\,\zeta_6\,\zeta_3^2
-\mbox{$\frac{13177\times15991}{691}$}\,\zeta_{12}\nonumber\\&&{}
+2^4\cdot3^3\cdot5\cdot7\,\zeta_{6,2}\,\zeta_4
-2^7\cdot3^3\,\zeta_{8,2}\,\zeta_2
-2^6\cdot3^2\cdot11^2\zeta_{10,2}\label{push}
\end{eqnarray}
which shows that, at weight 12, a depth 4 MZV is reducible to terms of
depth $d\le2$, and their products, if one allows an {\em alternating} double sum in the basis.

When constructing the MZV data mine, Bl\"{u}mlein, Broadhurst and Vermaseren~\cite{BBV}
proved this, by massive use of computer algebra.
It would be much harder to prove an LLL discovery
at weight 21 and depth 7, where
\begin{equation}81\zeta_{6,2,3,3,5,1,1}+326\sum_{j>k>l>m>n>0}\frac{(-1)^{k+m}}{(jk^2lm^2n)^3}
\label{depth4}\end{equation}
was reduced to 150 terms containing MZVs of depths $d\le5$. 

\section{Counterterms at 7 loops }

Broadhurst found reductions to MZVs for a pair of 7-loop periods~\cite{BS}
\begin{eqnarray}
P_{7,8}&=&
\frac{22383}{20}\zeta_{11}+\frac{4572}{5}\left(\zeta_{3,5,3}-\zeta_{3}\zeta_{5,3}\right)-700\zeta_{3}^2\zeta_{5}
\nonumber\\
&&\quad+\,1792\zeta_{3}\left(\frac{9}{320}\left(12\zeta_{5,3}-29\zeta_{8}\right)+\frac{45}{64}\zeta_{5}\zeta_{3}\right)
\label{P78}\\
P_{7,9}&=&
\frac{92943}{160}\zeta_{11}+\frac{3381}{20}\left(\zeta_{3,5,3}-\zeta_{3}\zeta_{5,3}\right)-\frac{1155}{4}\zeta_{3}^2\zeta_{5}\nonumber\\
&&\quad+\,896\zeta_{3}\left(\frac{9}{320}\left(12\zeta_{5,3}-29\zeta_{8}\right)+\frac{45}{64}\zeta_{5}\zeta_{3}\right)
\label{P79}
\end{eqnarray}
that had been expected to involve alternating sums.

These results were later proven, one by the methods of Erik Panzer~\cite{EP} and the other 
by the methods of Oliver Schnetz~\cite{OS}.  Their methods yielded complicated combinations of {\em alternating}
sums,  which were then reduced to the MZV formulas~(\ref{P78},\ref{P79}) 
by use of proven results in the MZV data mine~\cite{BBV}.

The period $P_{7,11}$ in the census of Schnetz~\cite{census} is much more demanding.
All other periods up to 7 loops reduce to MZVs; 
only $P_{7,11}$ requires multiple polylogarithms of {\em sixth} roots of unity,
of the form~(\ref{nest}) with $\sigma_j^6=1$.

Panzer evaluated $\sqrt{3}P_{7,11}$
in terms of 4589 such sums, each of which he evaluated to 5000 digits.
Then he found  an empirical reduction to
a 72-dimensional basis. The rational coefficient of
$\pi^{11}$ in his result was~\cite{EP}
\begin{equation}C_{11}=-\frac{964259961464176555529722140887}
{2733669078108291387021448260000}
\label{C11}\end{equation}
whose denominator contains 8 primes greater than 11,
namely 19, 31, 37, 43, 71, 73, 50909 and 121577.

Using LLL, it was possible to find a much better basis, 
with no prime greater 3 in the denominator of any coefficient.
Let $A={\rm d}\log(x)$, $B=-{\rm d}\log(1-x)$ and $D=-{\rm d}\log(1-\exp(2\pi {\rm i}/6)x)$
be letters, forming words $W$ that define iterated integrals $Z(W)$. Let
\begin{equation}W_{m,n}=\sum_{k=0}^{n-1}
\frac{\zeta_3^k}{k!}A^{m-2k}D^{n-k},\label{W}\end{equation}
$P_n=(\pi/3)^n/n!$, 
$I_n={\rm Cl}_n(2\pi/3)$ and
$I_{a,b}=\Im Z(A^{b-a-1}D A^{2a-1}B)$.
Using
\begin{eqnarray}
I_{2,9}&=&91(11T_{2,9}) - 898T_{3,8}
+ 11I_{4,7} - 292P_{11}\label{I29}\\
I_{3,8}&=&24(11T_{2,9}) + 841T_{3,8}
- 190I_{4,7} - 255P_{11}\label{I38}
\end{eqnarray}
to transform  to $T_{2,9}$ and $T_{3,8}$,
the result becomes
\begin{eqnarray}
\sqrt{3}P_{7,11}&=&
-10080\Im Z(W_{7,4}+W_{7,2}P_2)+50400\zeta_3\zeta_5P_3\nonumber\\&&{}
+\left(35280\Re Z(W_{8,2})+\frac{46130}{9}\zeta_3\zeta_7+17640\zeta_5^2\right)P_1\nonumber\\&&{}
-13277952T_{2,9}-7799049T_{3,8}+\frac{6765337}{2}I_{4,7}-\frac{583765}{6}I_{5,6}\nonumber\\&&{}
-\frac{121905}{4}\zeta_3I_8-93555\zeta_5I_6-102060\zeta_7I_4-141120\zeta_9I_2\nonumber\\&&{}
+\frac{42452687872649}{6}P_{11}.\label{p5}
\end{eqnarray}

\section{Periods and quasi-periods in electrodynamics}

The magnetic moment of the electron, in Bohr magnetons, has quantum  electrodynamic contributions
$\sum_{L=0}^4a_L(\alpha/\pi)^L$ given up to $L=4$ loops by~\cite{SL}
\begin{eqnarray}
a_0&=&\phantom{-}1\phantom{.5}\quad[{\rm Dirac},\,1928]\\
a_1&=&\phantom{-}0.5\quad[{\rm Schwinger},\,1947]\\
a_2&=&-0.32847896557919378458217281696489239241111929867962\ldots\\
a_3&=&\phantom{-}1.18124145658720000627475398221287785336878939093213\ldots\\
a_4&=&-1.91224576492644557415264716743983005406087339065872\ldots
\end{eqnarray}
In 1957, corrections by Petermann and by Sommerfield resulted  in
\begin{equation}a_2=\frac{197}{144}+\frac{\zeta_2}{2}+\frac{3\zeta_3-2\pi^2\log2}{4}\,.
\label{a2}\end{equation}
In 1996,  Laporta and Remiddi obtained
\begin{eqnarray}
a_3&=&\frac{28259}{5184}+\frac{17101\zeta_2}{135}+\frac{139\zeta_3-596\pi^2\log2}{18}\nonumber\\&&{}
-\frac{39\zeta_4+400U_{3,1}}{24}-\frac{215\zeta_5-166\zeta_3\zeta_2}{24}\,.\label{a3}
\end{eqnarray}
The 3-loop contribution contains a weight-4 depth-2 alternating sum
\begin{equation}U_{3,1}=\sum_{m>n>0}\frac{(-1)^{m+n}}{m^3n}=
\frac{\zeta_4}{2}+\frac{(\pi^2-\log^22)\log^22}{12}-2\sum_{n>0}\frac{1}{2^n n^4}.
\label{U31}\end{equation}

Equally fascinating is the Bessel moment~\cite{BR}
\begin{equation}B=-\int_0^\infty\frac{27550138t+35725423t^3}{48600}I_0(t)K_0^5(t){\rm d}t
\label{B}\end{equation}
in the evaluation by Laporta~\cite{SL}, at 4800 digits, of
\begin{equation}a_4=P+B+E+U\approx
2650.565-1483.685-1036.765-132.027\approx-1.912\label{comb}\end{equation}
where $P$ comprises multiple polylogs, $E$ comprises integrals
whose integrands contain logs and products of elliptic integrals
and $U$ comes from 6 light-by-light integrals, still under investigation.

\subsection{Bessel moments and modular forms}

Gauss noted on 30 May 1799 that the lemniscate constant
\begin{equation}\int_0^1\frac{{\rm d}x}{\sqrt{1-x^4}}=\frac{(\Gamma(1/4))^2}{4\sqrt{2\pi}}
=\frac{\pi/2}{{\rm agm}(1,\sqrt2)}\label{Gauss}\end{equation}
is given by the reciprocal of an arithmetic-geometric mean. 
This is an example of the Chowla-Selberg formula at the first singular value~\cite{piagm}.
In 1939, Watson~\cite{GNW} encountered the sixth singular value, in work on integrals from condensed matter physics.
Here, $\left(\sum_{n\in{\bf Z}}\exp(-\sqrt{6}\pi n^2)\right)^4$
gives the product of $\Gamma(k/{24})$ with $k=1,5,7,11$, as observed by Glasser and Zucker~\cite{GZ} in 1977.
In 2007, Broadhurst and Laporta identified a Feynman period at the {\em fifteenth} singular value~\cite{BBBG}, 
where $\left(\sum_{n\in{\bf Z}}\exp(-\sqrt{15}\pi n^2)\right)^4$
gives the product of $\Gamma(k/{15})$ with $k=1,2,4,8$.

With $N=a+b$ Bessel functions and $c\geq0$, we define moments
\begin{equation}M(a,b,c)= \int_0^\infty I_0^a(t)K_0^b(t)t^c{\rm d}t\label{Mabc}\end{equation}
that converge for $b>a\ge0$. Then the 5-Bessel matrix
\begin{equation}\left[\begin{array}{lr}M(1,4,1)&M(1,4,3)\\M(2,3,1)&M(2,3,3)\end{array}\right]\,=\,
\left[\begin{array}{lr}
\pi^2C&\pi^2\left(\frac{2}{15}\right)^2\left(13C-\frac{1}{10C}\right)\\
\frac{\sqrt{15}\pi}{2}C&\frac{\sqrt{15}\pi}{2}\left(\frac{2}{15}\right)^2\left(13C+\frac{1}{10C}\right)
\end{array}\right]\label{matC}\end{equation}
involves a single new constant
\begin{equation}C=\frac{\pi}{16}\left(1-\frac{1}{\sqrt{5}}\right)
\left(\sum_{n=-\infty}^\infty\exp(-\sqrt{15}\pi n^2)\right)^4
=\frac{1}{240\sqrt5\pi^2}\prod_{k=0}^3
\Gamma\left(\frac{2^k}{15}\right)\label{C}\end{equation}
and its {\em reciprocal}. The determinant $2\pi^3/\sqrt{3^3 5^5}$ of matrix~(\ref{matC}) is an algebraic
multiple of a power of  $\pi$. This is an example of an all-loop result discovered by Broadhurst
and Mellit~\cite{BM} and proven by Yajun Zhou~\cite{Zhou1}.  

The L-series for $N=5$ Bessel functions comes from a {\em modular form}
of weight 3 and level 15:
\begin{eqnarray}
\eta_n&= &q^{n/24}\prod_{k>0}(1-q^{nk}),\quad q=\exp(2\pi{\rm i}\tau),\label{eta}\\
f_{3,15}(\tau)&= &(\eta_3\eta_5)^3+(\eta_1\eta_{15})^3=\sum_{n>0}A_5(n) q^n\label{f315}\\
L_5(s)&= &\sum_{n>0}\frac{A_5(n)}{n^s}\quad{\rm for}~s>2\label{L5s}\\
L_5(1)&=&\sum_{n>0}\frac{A_5(n)}{n}\left(2+\frac{\sqrt{15}}{2\pi n}\right)\exp\left(-\frac{2\pi n}{\sqrt{15}}\right)
\label{L51}\\
&=&5C\,=\,\frac{5}{\pi^2}\int_0^\infty I_0(t)K_0^4(t)t{\rm d}t\,.\label{5C}
\end{eqnarray}

\subsection{Periods and quasi-periods for the Laporta problem}

Laporta's 4-loop work~\cite{SL} engages the first row of the 6-Bessel determinant
\begin{equation}
{\rm det}\left[\begin{array}{lr}M(1,5,1)&M(1,5,3)\\M(2,4,1)&M(2,4,3)\end{array}\right]=\frac{5\zeta_4}{32}
\label{Mat6}\end{equation}
associated to a modular form $ f_{4,6}(\tau)=(\eta_1\eta_2\eta_3\eta_6)^2$
with weight 4 and level 6~\cite{BS,Zhou2}.  At top left we have $M(1,5,1)$, from the on-shell
4-loop sunrise diagram, in two spacetime dimensions. Below it, $M(2,4,1)$ comes from cutting
an internal line. The second column comes from differentiating the first, with respect to the
external momentum, to produce quasi-periods associated with a {\em weakly} holomorphic
modular form
\begin{equation}
\widehat{f}_{4,6}(\tau)=\mu{f}_{4,6}(\tau),\quad
\mu=\frac{1}{32}\left(w+\frac3w\right)^4-\frac{9}{16}\left(w+\frac3w\right)^2,\quad
w=\frac{3\eta_3^4\eta_2^2}{\eta_1^4\eta_6^2}.\label{f46}
\end{equation}
With $s=1,2$, we computed compute 10,000 digits of the Eichler integrals
\begin{equation}
\frac{\Omega_s}{(2\pi)^s}=\int_{1/\sqrt{3}}^\infty{f}_{4,6}\left(\frac{1+{\rm i}y}2\right)y^{s-1}dy,\quad
\frac{\widehat\Omega_s}{(2\pi)^s}=
\int_{1/\sqrt{3}}^\infty\widehat{f}_{4,6}\left(\frac{1+{\rm i}y}2\right)y^{s-1}dy\label{ohat}.
\end{equation}

\subsection{Laporta's intersection number}
LLL readily gave 4 linear relations
\begin{equation}
\frac{2}{\pi^2}\left[\begin{array}{rr}
4M(1,5,1)&\mbox{$\frac{36}{5}$}\left(M(1,5,1)+M(1,5,3)\right)\\[5pt]
\mbox{$\frac{5}{3}$}M(2,4,1)& 3\left(M(2,4,1)+M(2,4,3)\right)
\end{array}\right]=
\left[\begin{array}{rr}
-\Omega_2&\widehat{\Omega}_2\\
-\Omega_1& \widehat{\Omega}_1
\end{array}\right]
\label{matrel}
\end{equation}
between Feynman integrals, the periods $\Omega_{1,2}$ and the quasi-periods $\widehat{\Omega}_{1,2}$.

The intersection number is the determinant of this matrix, namely $1/12$. 
Broadhurst and Roberts~\cite{BR} converted this into a quadratic relation between 4 hypergeometric series:
\begin{equation}\begin{array}{lclrrrrrrrrr}
F_a&=&{}_4F_3(& 1/2,&2/3, &2/3, &5/6;&7/6, &7/6,& 4/3;&1)\\
F_b&=&{}_4F_3(&-1/2,&1/6, &1/3, &4/3;&-1/6,& 5/6,& 5/3;&1)\\
F_c&=&{}_4F_3(& 1/6,& 1/3,& 1/3, &1/2;&2/3,& 5/6,& 5/6;&1)\\
F_d&=&{}_4F_3(&-7/6,&-1/2,&-1/3,& 2/3;&-5/6,& 1/6,& 1/3;&1)
\end{array}\label{hyp}\end{equation}
namely
\begin{equation}7F_aF_b+10F_cF_d=40,\label{matF}\end{equation}
which was later proven by Yajun Zhou~\cite{Zhou-Lap}.

\section{Quadratic relations}

In this section, we give quadratic relations between Feynman integrals,
recently discovered by using the LLL algorithm. If one has $n$ integrals,
there are $n(n+1)/2$ products to consider, in linear combinations
with rational or algebraic coefficients,  which may give a rational or algebraic
multiple of a power of $\pi$. This problem soon explodes. We begin with
a conjecture obtained after intensive use of LLL and tested with up to
$n=100$ Feynman integrals.

\noindent{\bf Conjecture:} (Broadhurst and Roberts~\cite{BR})
 {\em With the Feynman, de Rham and Betti matrices below,}
\begin{equation}F_N^{} D_N^{} F_N^{\tt tr}=B_N^{}.\label{conj}\end{equation}

The elements of the Feynman matrices $F_N$ are the Bessel moments
\begin{eqnarray}
F_{2k+1}(u,a)&=&\frac{(-1)^{a-1}}{\pi^u}M(k+1-u,k+u,2a-1)\label{Fo}\\
F_{2k+2}(u,a)&=&\frac{(-1)^{a-1}}{\pi^{u+1/2}}M(k+1-u,k+1+u,2a-1)\label{Fe}
\end{eqnarray}
with $u$ and $a$, as well as later indices $v$ and $b$, running from 1 to $k$.
$F_N^{\tt tr}$ is the transpose of $F_N$.

The Betti matrices $B_N$ have rational elements given by 
\begin{eqnarray}
B_{2k+1}(u,v)&=&(-1)^{u+k}2^{-2k-2}(k+u)!(k+v)!Z(u+v)\label{Bo}\\
B_{2k+2}(u,v)&=&(-1)^{u+k}2^{-2k-3}(k+u+1)!(k+v+1)!Z(u+v+1)\label{Be}\\
Z(m)&=&\frac{1+(-1)^m}{(2\pi)^m}\zeta_m.\label{Zm}
\end{eqnarray}

For the de Rham matrices $D_N$, let
$v_k$ and $w_k$ be the rationals generated by
\begin{eqnarray}\frac{J_0^2(t)}{C(t)}&=&\sum_{k\geq0}\frac{v_k}{k!}\left(\frac{t}{2}\right)^{2k}=
1-\frac{17t^2}{54}+\frac{3781t^4}{186624}+\ldots\label{vex}\\
\frac{2J_0(t)J_1(t)}{tC(t)}&=&\sum_{k\geq0}\frac{w_k}{k!}\left(\frac{t}{2}\right)^{2k}=
1-\frac{41t^2}{216}+\frac{325t^4}{186624}+\ldots\label{wex}\end{eqnarray}
where $J_0(t)=I_0({\rm i}t)$, $J_1(t)=-J_0^\prime(t)$ and
\begin{equation}C(t)= \frac{32(1-J_0^2(t)-t J_0(t)J_1(t))}{3t^4}=
1-\frac{5t^2}{27}+\frac{35t^4}{2304}-\frac{7t^6}{9600}+\ldots\label{Cex}\end{equation}
Construct rational bivariate polynomials $H_s = H_s(y,z)$ by the recursion 
\begin{equation}H_s=zH_{s-1}-(s-1)y H_{s-2}-\sum_{k=1}^{s-1} 
{s-1\choose k}\left(v_k H_{s-k}-w_k z H_{s-k-1}\right)\label{Hpol}\end{equation} 
for $s>0$, with $H_0=1$ and $H_{-1}=0$. 
Use these to define 
\begin{equation}d_s(N,c)= \frac{H_s(3c/2,N+2-2c)}{4^s s!}.\label{dpol}\end{equation} 
Finally, construct de Rham matrices with the rational  elements 
\begin{equation}D_N(a,b)= \sum_{c=-b}^{a}d_{a-c}(N,-c)d_{b+c}(N,c)c^{N+1}.\label{dR}\end{equation} 

The discovery of formula~(\ref{dR}) for the coefficients of these quadratic relations
involved intensive use of LLL, at high numerical precision. At 20 loops,
there are 100 Feynman integrals to consider,  with 5050 products.   
Javier Fres\'an, Claude Sabbah and Jeng-Daw Yu~\cite{FSY}
have verified that our formulas hold up to 20 loops, after which they ran out  of computing power.
They encountered subtleties when $N$ is divisible by 4. These are entirely avoided by our uniform formula~(\ref{dR}).
Yajun Zhou~\cite{Zhou3} has given an illuminating classical proof, with generalizations.

\subsection{Quadratic relations at weight 6 and level 24}

Here we establish relations between Bessel moments~(\ref{Mabc})
with  $a+b=6$, $c=0,2,4$, and Eichler integrals of
modular forms of weight 6 and level 24, constructed from eta quotients.
This connection was suggested by the discovery of the linear and quadratic relations
\begin{eqnarray}
\frac{M(0,6,0)}{M(2,4,0)}=\frac{3M(0,6,2)-8M(0,6,4)}{3M(2,4,2)-8M(2,4,4)}&=&3\pi^2,\label{Mlin}\\
{\rm det}\left[\begin{array}{lr}M(0,6,0)&3M(0,6,2)-8M(0,6,4)\\
M(1,5,0)&3M(1,5,2)-8M(1,5,4)\end{array}\right]&=&\frac{5\pi^6}{16}.\label{Mquad}
\end{eqnarray}

We begin by defining three eta quotients, subject to two algebraic relations:
\begin{eqnarray}r&=&\left(\frac{\eta_{2}\eta_{12}}{\eta_{4}\eta_{6}}\right)^{6}
=\frac{s-t}{st}=9t-8s
=q - 6q^3 + 15q^5 + O(q^7),\label{ris}\\
s&=&\left(\frac{\eta_{4}\eta_{12}}{\eta_{2}\eta_{6}}\right)^{3}
=q + 3q^3 + 6q^5 + O(q^7),\label{sis}\\
t&=&\left(\frac{\eta_{6}\eta_{12}}{\eta_{2}\eta_{4}}\right)^{2}
=q + 2q^3 + 7q^5 + O(q^7).\label{tis}\end{eqnarray}
The moments $M(0,6,0)$ and $M(1,5,0)$ are related to periods of the modular form
\begin{equation}f_1(\tau)=\left(\eta_{2}\eta_{4}\eta_{6}\eta_{12}\right)^3\left(\frac{1}{s^2}-64s^2\right)
=q-9q^3-34q^5-240q^7+81q^9+O(q^{11}).\label{f1}\end{equation}
To find its quasi-periods, we form a column vector of 5 cusp forms
\begin{eqnarray}{\bf f}(\tau)&=&\left(\eta_{2}\eta_{4}\eta_{6}\eta_{12}\right)^3
\left[\begin{array}{lllll}
1/s^2-64s^2\\1/s^2+64s^2\\1/r^2-r^2\\1/t^2-81t^2\\1/t^2+81t^2+54\end{array}\right]
={\bf T}\left[\begin{array}{lllll}q\\q^3\\q^5\\q^7\\q^9\end{array}\right]+O(q^{11}),\label{fex}\\
{\bf T}&=&\left[\begin{array}{rrrrr}
1& -9& -34& -240&   81\\
1& -9&  94&  144&   81\\
1&  9&  38&  120&   81\\
1& -7& -74&  -24& -383\\
1& 47& -74&  -24&  697\end{array}\right],\label{matT}\end{eqnarray}
with the Hecke matrix ${\bf T}$ recording the first 5 non-vanishing Fourier coefficients of
the 5 modular forms. The first three components of ${\bf f}$ are new forms, while the remaining
two are old forms.

Since there are 16 cusp forms of weight 6 and level 24, we began our investigation
with a far more fearsome 16-dimensional problem. After intensive study of the relationship between
Rademacher sums~\cite{AB} and the determinants and permanents~\cite{perm} of matrices of periods and quasi-periods
of modular forms, we were able to reduce the Bessel-moment problem
to the 5-dimensional problem presented here.

For each of the 5 cusp forms $f_k$, we seek a weakly holomorphic form $\widehat{f}_k$,
such that the periods of $f_k$ and quasi-periods of $\widehat{f}_k$ yield a determinant
that is a rational multiple of a power of $\pi$. For $k=1$, this will solve the Bessel-moment problem.
 
To construct 5 weakly homomorphic modular forms, we define a column vector
\begin{equation}{\bf g(\tau)}=\left(\frac{\eta_{12}^5}{\eta_{4}^{}\eta_{6}^2}\right)^6
\left[\begin{array}{l}
1\\
a   + 35\\
a^2 + 40a   + 646\\
a^3 + 45a^2 + 840a    + 8352\\
a^4 + 50a^3 + 1059a^2 + 12308a + 84817\end{array}\right],\label{gtau}\,\quad
a=72\frac{\eta_4^{}\eta_{12}^5}{\eta_2^5\eta_6^{}},\label{gpol}
\end{equation}
with monic polynomials in $a=O(q^2)$, determined by the requirement that
\begin{equation}\frac{-27}{\tau^6}g_k\left(-\frac{1}{24\tau}\right)=\frac{1}{q^{2k-1}}+O(q),\label{sing}\end{equation}
which records the singular behaviour near the cusp of $g_k(\tau)$ at $\tau=0$.
We avoid this singularity by taking Eichler integrals from $\tau=\frac14$ to
$\tau={\rm i}\infty$, with extremely good behaviour of  $g_k(\tau)$ at the end-points.

The Fourier expansion of $g_k$ begins at $q^{11}$. Thus we may add, to any 
combination of the weakly holomorphic forms in ${\bf g}$, a combination
of the cusp forms in ${\bf f}$, since the latter are determined by their expansions
up to $q^9$, recorded in the Hecke matrix ${\bf T}$.  Our Ansatz for the
weakly homomorphic partners in $\widehat{\bf f}$
has the form
\begin{equation}\widehat{\bf f}=\widetilde{\bf T}^{-1}({\bf Ug+VT^{\rm-1}f}),\quad
{\bf U}=18^2\left[\begin{array}{ccccc}
1&    0&       0&        0&          0\\
0&    3^5&       0&        0&          0\\
0&    0&       5^5&        0&          0\\
0&    0&       0&        7^5&          0\\
0&    0&       0&        0&          9^5\end{array}\right],\label{matU}\end{equation}
which uses inverses of ${\bf T}$ and its transpose $\widetilde{\bf T}$.
The diagonal matrix ${\bf U}$ reflects the singular
behaviour~(\ref{sing}) of the weakly holomorphic modular forms $g_k$ near $\tau=0$.

Our final challenge is to determine the matrix ${\bf V}$, which we expect,
from previous work, to be symmetric, with a vanishing first row and column.
Here we give our eventual result
\begin{equation}{\bf V}=6^2\left[\begin{array}{rrrrr}
0&    0&       0&        0&          0\\
0&   -4&      54&      648&       5995\\
0&   54&    2916&    77508&    1150848\\
0&  648&   77508&  3039444&   64431936\\
0& 5995& 1150848& 64431936& 1865595908\end{array}\right],\label{matV}\end{equation}
with explanation of how it was obtained, empirically, by integer-relation searches.

We define periods, quasi-periods and determinants as follows: 
\begin{equation}\left[\begin{array}{c}P_k(s)\\\widehat{P}_k(s)\end{array}\right]
=-{\rm i}\int_0^\infty
\left[\begin{array}{c}f_k((1+{\rm i}y)/4)\\\widehat{f}_k((1+{\rm i}y)/4)\end{array}\right]
y^{s-1}{\rm d}y,\quad D_k(s,t)=P_k(s)\widehat{P}_k(t)-P_k(t)\widehat{P}_k(s),\label{pphat}\end{equation}
with $k=1,2,3,4,5$ and $s=1,2,3,4,5$. The Eichler integrals are from $\tau=\frac14$ to $\tau={\rm i}\infty$,
along the vertical line where the real part of $\tau$ is $\frac14$ and hence $q=\exp(2\pi{\rm i}\tau)$
is pure imaginary. Since $f_k$ and $\widehat{f}_k$ have Fourier expansions in odd powers of $q$,
they too are pure imaginary. Hence the periods and quasi-periods are real. 

Our criteria for the elements of ${\bf V}$ are the determinant conditions
\begin{equation}D_k(1,3)=D_k(1,5)=D_k(2,4)=0,\quad D_k(1,2)=\frac{d_k}{\pi^5},\label{cond}\end{equation}
where $d_k$ is a rational number. The first three conditions ensure the matching
of the period polynomial of $f_k$ with the quasi-period polynomial of $\widehat{f}_k$. The fourth
condition encodes a non-trivial quadratic relation between periods and quasi-periods.
Moreover, we were able to use Rademacher sums~\cite{AB} and Petersson inner products~\cite{GP}
to show that $d_k=\frac{21}{2},-\frac{135}{14},-\frac{15}{2},\frac{15}{2},-\frac{6}{7}$,
for $k=1,2,3,4,5$.

Thus we have $5\times4=20$ integer-relation conditions with which to determine
$4+3+2+1=10$ independent elements of ${\bf V}$, giving us great confidence
in our results.

Here we record the relations between periods, quasi-periods and Bessel moments at $k=1$,
which have been checked at 1000-digit precision:
\begin{eqnarray}
P_1(1)=21P_1(3)=9P_1(5)&=&28\,\frac{M(0,6,0)}{\pi^6},\label{P1}\\
P_1(2)=3P_1(4)&=&8\,\frac{M(1,5,0)}{\pi^5},\label{P2h}\\
\widehat{P}_1(1)=21\widehat{P}_1(3)=9\widehat{P}_1(5)&=&7\,
\frac{17M(0,6,0)+48(3M(0,6,2)-8M(0,6,4))}{80\pi^6},\label{Ph1}\\
\widehat{P}_1(2)=3\widehat{P}_1(4)&=&2\,
\frac{17M(1,5,0)+48(3M(1,5,2)-8M(1,5,4))}{80\pi^5}.\label{Ph2}
\end{eqnarray}

\subsection{Quadratic relations at levels 14 and 34}

In  2019, Philip Candelas, Xenia de la Ossa, Mohamed Elmi and Duco van Straten
announced a discovery of a family of Calabi-Yau manifolds with rank-2 attractor points~\cite{Cand}.
 
They compactified  a 10-dimensional supergravity theory on a 
Calabi-Yau three-fold with complex structure, to obtain 4-dimensional black holes, with event horizons
whose areas are determined by their electric and magnetic charges and by ratios of periods
of modular forms of weight 4 and levels 14 or 34. 

Hearing of this on a visit to Oxford in November 2019, Broadhurst
observed that their Calabi-Yau periods come from solutions to a homogeneous differential equation
associated with 4-loop sunrise integrals, namely
\begin{eqnarray}
M_{m,n}(z)&=&\int_0^\infty I_0(xz)[I_0(x)]^m[K_0(x)]^{5-m}x^{2n+1}{\rm d}x\label{Mmn}\\
N_{m,n}(z)&=&z\int_0^\infty I_1(xz)[I_0(x)]^m[K_0(x)]^{5-m}x^{2n+2}{\rm d}x\label{Nmn}
\end{eqnarray}
with $m\in\{0,1,2\}$, integers $n\ge0$ and real $z^2<(5-2m)^2$. The uncut
diagram $M_{0,0}(z)$ satisfies an {\em inhomogeneous} differential equation.

The external mass is $z$. At $z=1$ we obtain Laporta's on-shell periods, for the magnetic moment
of the electron at 4 loops,
coming from the modular form $f_{4,6}(\tau)=(\eta_1\eta_2\eta_3\eta_6)^2$ with level 6.
With mass $z=\sqrt{17}-4$, we obtain level-34 periods. At the space-like point
$z=\sqrt{-7}$, we obtain level-14 periods.

At each of the levels 14 and 34, Candelas et al.\ considered 16 Calabi-Yau periods,
coming from 4 solutions to a homogeneous fourth-order differential equation, together with
the first 3 derivatives of each solution. They were unable to identify all of these 16 periods.

Using LLL we found that 8 Feynman integrals, at each level, suffice to solve their problem, completely.
These  8 integrals determine a pair of periods and a pair of quasi-periods, at each of the weights
2 and 4. Hence they satisfy two quadratic relations. At level 34, the coefficients in these
relations are algebraic numbers in ${\bf Q}(\sqrt{17})$.

\subsection{Level 14, at space-like momentum}

At level 14, with $z=\sqrt{-7}$, we identified
\begin{equation}
f_{4,14}(\tau)=
 \frac{(\eta_2\eta_7)^6}{(\eta_1\eta_{14})^2}-4(\eta_1\eta_2\eta_7\eta_{14})^2
+\frac{(\eta_1\eta_{14})^6}{(\eta_2\eta_7)^2}\label{f414}
\end{equation}
as the relevant modular form of weight 4.
Its periods are critical values of the L-function
$L(f_{4,14},s)=((2\pi)^s/\Gamma(s))\int_0^\infty f_{4,14}({\rm i}y)y^{s-1}{\rm d}y$, with
\begin{eqnarray}
L(f_{4,14},3)&=&M_{1,0}(\sqrt{-7})
=\int_0^\infty J_0(\sqrt{7}x)I_0(x)K_0^4(x)x{\rm d}x=\frac{\pi^2}{7}L(f_{4,14},1)\label{easy}\\
\mbox{$\frac{1}{2}$}L(f_{4,14},2)&=&M_{2,0}(\sqrt{-7})
=\int_0^\infty J_0(\sqrt{7}x)I_0^2(x)K_0^3(x)x{\rm d}x. \label{next}
\end{eqnarray}

There is also a modular form of weight 2 to consider, $f_{2,14}(\tau)=\eta_1\eta_2\eta_7\eta_{14}$.
This provides a modular parametrization of a quartic elliptic curve, namely
\begin{eqnarray}
d^2&=&(1+h)(1+8h)(1+5h+8h^2),\label{curve}\\
h&=&\left(\frac{\eta_2\eta_{14}}{\eta_1\eta_7}\right)^3=q + 3q^2 + 6q^3 + 13q^4 + O(q^5),\label{his}\\
d&=&\frac{q}{f_{2,14}}\frac{{\rm d}h}{{\rm d}q}=1 + 7q + 27q^2 + 92q^3 + 259q^4 + O(q^5),\label{dis}
\end{eqnarray}
yielding an  L-value and a j-invariant:
\begin{equation} 
L(f_{2,14},1)=\frac{\omega_+}{3},\quad
j\left(\frac{\omega_++{\rm i}\omega_-}{2\omega_+}\right)=\left(\frac{5\times43}{28}\right)^3,\label{jis}
\end{equation}
with periods determined by arithmetic-geometric means
\begin{equation}
\omega_{\pm}=\frac{2\pi}{{\rm agm}\left(\sqrt{2^{9/2}\pm13},\,2^{11/4}\right)}\label{agm}
\end{equation}
and also by Feynman integrals:
\begin{eqnarray}
\frac{  \omega_+}{2}&=&3M_{2,0}(\sqrt{-7}) + 4N_{2,0}(\sqrt{-7}),\label{omp}\\
\frac{\pi\omega_-}{2}&=&3M_{1,0}(\sqrt{-7}) + 4N_{1,0}(\sqrt{-7}).\label{omm} 
\end{eqnarray}

The quasi-periods at weight 2 are $\widehat{\omega}_\pm$, with
\begin{eqnarray}
\frac{3  \widehat{\omega}_+}{16}&=&7M_{2,0}(\sqrt{-7}) + 8N_{2,0}(\sqrt{-7})+28M_{2,1}(\sqrt{-7}),\label{omhp}\\
\frac{3\pi\widehat{\omega}_-}{16}&=&7M_{1,0}(\sqrt{-7})+ 8N_{1,0}(\sqrt{-7})+28M_{1,1}(\sqrt{-7}) .\label{omhm} 
\end{eqnarray}
Suppressing the argument $z=\sqrt{-7}$, we obtain the quadratic relation
\begin{equation}
{\rm det}\left[\begin{array}{lr}
3M_{2,0}+4N_{2,0}&M_{2,0}+28M_{2,1}\\
3M_{1,0}+4N_{1,0}&M_{1,0}+28M_{1,1}
\end{array}\right]=-\frac{3\pi^2}{32}\label{det214}
\end{equation}
from Legendre's relation for complete elliptic integrals. 

At weight 4 we found
\begin{equation}
{\rm det}\left[\begin{array}{lr}
M_{2,0}&39N_{2,0}-427M_{2,1}-112N_{2,1}\\
M_{1,0}&39N_{1,0}-427M_{1,1}-112N_{1,1}
\end{array}\right]=\frac{3\pi^2}{32}\label{det414}
\end{equation}
as the quadratic relation between the periods and quasi-periods of $f_{4,14}$.

We define the weight-4 periods and quasi-periods as 
\begin{equation}{\cal G}_m=M_{m,0},\quad\widehat{\cal G}_m= 
7(35M_{m,0}-122M_{m,1})+2(39N_{m,0}-112N_{m,1}),\label{Ghat}
\end{equation}
for $m=1,2$ and $z=\sqrt{-7}$.
The quasi-periods come from Eichler integrals of a weakly holomorphic form
obtained by multiplying $f_{2,14}^2$ by a polynomial that is quartic in
$h$ and linear in $d/h$.
For one of the quasi-periods,
the dependence on $d/h$ is irrelevant. We used LLL to determine that
\begin{eqnarray}
\widehat{\cal G}_2&=&5\pi^2\int_{1/\sqrt7}^\infty g\left(\frac{1+{\rm i}y}{2}\right)y{\rm d}y\\
\quad g(\tau)&=&(253+645h+1446h^2+2064h^3+1024h^4)f_{2,14}^2.
\end{eqnarray}
Then the other quasi-period
comes from the determinant ${\cal G}_2\widehat{\cal G}_1-{\cal G}_1\widehat{\cal G}_2=3(\pi/4)^2$. 

\subsection{Level 34, with mass $\sqrt{17}-4$}

At level 34, with $z=u=\sqrt{17}-4$, we used Pari-GP to identify the modular form of weight 4.
Let $\chi(n)$ be the Dirichlet character defined for prime $p$ by
$\chi(17)=0$ and otherwise by $\chi(p)=\pm1$ according as whether $p$ is or is not
a square modulo 17. Pari-GP declares that there are 12 cusp forms
of level 34 and weight 4 with this character.
Feynman integrals choose a pair of new forms whose Fourier
coefficients, $A_4(n)$ and $\overline{A}_4(n)$, are Gaussian integers,
related by complex conjugation. 

Let $L_4(s)$ be the analytic continuation of
\begin{equation}L_4(s)=\sum_{n>0}\frac{A_4(n)}{n^s}=\frac{1}{1+2^{1-s}}
\prod_{p>2}\frac{1}{1-A_4(p)p^{-s}+\chi(p)p^{3-2s}}\label{L4}
\end{equation}
with the choice of sign $A_4(3)=2{\rm i}$. For prime $p$, $A_4(p)$ is real if $\chi(p)=+1$ 
and imaginary if $\chi(p)=-1$, while $A_4(17)/17=1-4{\rm i}$ is truly complex. 

Feynman integrals determine the critical L-values at weight 4:
\begin{eqnarray}L_4(3)&=&\left(\frac{13-u+(1+13u){\rm i}}{17}\right)M_{1,0}(u),\label{L3}\\
L_4(2)&=&4\left(\frac{5-3u+(3+5u){\rm i}}{17}\right)M_{2,0}(u),\label{L2}\\
L_4(1)&=&\left(\frac{7-11u+(11+7u){\rm i}}{\pi^2}\right)M_{1,0}(u).\label{L1M}
\end{eqnarray}

At weight 2 they determine the periods and quasi-periods of the elliptic curve
 \begin{equation}
y^2=\left(x+\frac{5-u}{8}\right)\left(x+\frac{5+u}{8}\right)\left(x+\frac{3+u}{2}\right)\label{cubic}
\end{equation}
whose real and imaginary periods are
\begin{equation}
\omega_1=\frac{4\pi}{{\rm agm}\left(\sqrt{4u},\sqrt{14+10u}\right)},\quad
\omega_2=\frac{-4\pi{\rm i}}{{\rm agm}\left(\sqrt{14+6u},\sqrt{14+10u}\right)}.\label{omz}
\end{equation}

The elliptic periods $\omega_{1,2}$ and quasi-periods $\widehat{\omega}_{1,2}$ are determined by
\begin{eqnarray}
\frac{\omega_1}{4}&=&{\cal P}_2=(2+3u)M_{2,0}(u)+4(4+u)N_{2,0}(u)\label{om1}\\
\frac{\pi{\rm i}\omega_2}{4}&=&{\cal P}_1=(2+3u)M_{1,0}(u)+4(4+u)N_{1,0}(u)\label{om2}\\
\frac{3\widehat{\omega}_1}{8(1+u)}&=&
\widehat{\cal P}_2=M_{2,0}(u)+2(5+u)N_{2,0}(u)+2u(3+u)(4+u)M_{2,1}(u)\label{et1}\\
\frac{3\pi{\rm i}\widehat{\omega}_2}{8(1+u)}&=&
\widehat{\cal P}_1=M_{1,0}(u)+2(5+u)N_{1,0}(u)+2u(3+u)(4+u)M_{1,1}(u)\label{et2}
\end{eqnarray}
with Legendre's condition giving
${\cal P}_1\widehat{\cal P}_2-{\cal P}_2\widehat{\cal P}_1=3(\pi/4)^2/(1+u).$

At weight 4, the periods ${\cal H}_m=M_{m,0}(u)$  and quasi-periods
\begin{eqnarray}\widehat{\cal H}_m&=&81M_{m,0}(u)+3(2+u)(u-6)N_{m,0}(u)\nonumber\\
&+&{}u^2(2+u)(4+u)(96+11u)M_{m,1}(u)+136(1-u)N_{m,1}(u)
\end{eqnarray}
yield the intersection number
${\cal H}_1\widehat{\cal H}_2-{\cal H}_2\widehat{\cal H}_1=3(\pi/8)^2/u.$

The numbers which remained unidentified in~\cite{Cand} are now easy to determine.
They involve the permanents~\cite{perm} of the matrices of Feynman integrals whose determinants
yield intersection numbers that are algebraic multiples of powers of $\pi$.

\section{Summary}

PSLQ and LLL have enlivened quests for analytical results,  provided strong tests on conjectures
and condensed huge expressions. Parallel PSLQ was of the essence in Laporta's work in electrodynamics.
LLL led to a conjecture on quadratic relations for all loops,
to determinations of quasi-periods at weight 6 and level 24
and to exact results for black-hole problems that involve modular forms of levels 14 and 34.
Our new results at levels 14, 24 and 34 were obtained from extending methods developed in our work on 
eta quotients~\cite{AB}. The permanents which were lacking in~\cite{Cand}
yield Rademacher sums~\cite{AB} that are sums of products
of Bessel functions and Kloosterman sums~\cite{filskm} .

\raggedright

\end{document}